\documentclass[journal,10pt]{IEEEtran}
\usepackage[printonlyused]{acronym}
\usepackage[nobreak]{cite}
\usepackage{amsmath,amssymb,amsfonts}
\usepackage{bm}
\usepackage[dvipsnames]{xcolor}
\usepackage{graphicx}
\usepackage{booktabs}
\usepackage[capitalise]{cleveref}
\crefformat{equation}{#2(#1)#3}
\Crefformat{equation}{#2(#1)#3}
\usepackage{multirow}
\usepackage{subcaption}
\usepackage[utf8]{inputenc}
\usepackage{tikz}
\usetikzlibrary{calc}
\usepackage{url}

\newcommand{\positiontextbox}[4][]{%
  \begin{tikzpicture}[remember picture,overlay]
    \node[inner sep=3pt, fill=yellow,align=left,draw,line width=1pt,#1] at ($(current page.north west) + (#2,-#3)$) {\parbox{.95\paperwidth}{#4}};
  \end{tikzpicture}%
}

\begin{document}



\newacro{3GPP}[3GPP]{3rd Generation Partnership Project}
\newacro{5G}[5G]{fifth generation}
\newacro{6G}[6G]{sixth generation}
\newacro{AWGN}{additive white Gaussian noise}
\newacro{BER}{bit error ratio}
\newacro{BEP}{bit error probability}
\newacro{CI}{confidence interval}
\newacro{CP}{cyclic prefix}
\newacro{CDF}{cummulative density function}
\newacro{CIR}{channel impulse response}
\newacro{CLT}{central limit theorem}
\newacro{CQI}{channel quality indicator}
\newacro{CSI}{channel state information}
\newacro{DCT}{discrete cosine transform}
\newacro{EM}{expectation-maximization}
\newacro{FEC}{forward error correction}
\newacro{FFT}[FFT]{fast Fourier transform}
\newacro{ICI}{inter-carrier-interference}
\newacro{IDCT}{inverse discrete cosine transform}
\newacro{IDFT}{inverse discrete Fourier transform}
\newacro{IFFT}{inverse fast Fourier transform}
\newacro{ITU}{International Telecommunication Union}
\newacro{ISI}{inter-symbol-interference}
\newacro{JSCC}{joint source-channel coding}
\newacro{LMMSE}{linear minimum mean square error}
\newacro{LOS}{line-of-sight}
\newacro{LTE}{Long-Term Evolution}
\newacro{MCS}{modulation and coding scheme}
\newacro{MIMO}{multiple-input and multiple-output}
\newacro{MISO}{multiple-input and single-output}
\newacro{ML}{maximum likelihood}
\newacro{MMSE}{minimum mean squared error}
\newacro{MSE}{mean squared error}
\newacro{NLOS}{non-line-of-sight}
\newacro{OFDM}{orthogonal frequency-division multiplexing}
\newacro{PDF}{probability density function}
\newacro{PSK}{phase shift keying}
\newacro{PSNR}{peak signal-to-noise ratio}
\newacro{QAM}{quadrature amplitude modulation}
\newacro{SAGE}{space-alternating generalized \acl{EM}}
\newacro{SDR}{signal-to-distortion ratio}
\newacro{SEP}{symbol error probability}
\newacro{SIMO}{single-input and multiple-output}
\newacro{SINR}{signal-to-interference-plus-noise ratio}
\newacro{SISO}{single-input and single-output}
\newacro{SNR}{signal-to-noise ratio}
\newacro{UAV}[UAV]{unmanned aerial vehicles}
\newacro{V2I}{vehicle-to-infrastructure}
\newacro{V2V}{vehicle-to-vehicle}
\newacro{WSS}{wide-sense stationary}
\newacro{WSSUS}{WSS with uncorrelated scatterers}

\title{Bit Error Probability and Capacity Bound of OFDM Systems in Deterministic Doubly-Selective Channels}

\author{Tom\'as~Dom\'inguez-Bola\~no,~
	Jos\'e~Rodr\'iguez-Pi\~neiro,~ 
	Jos\'e A. Garc\'ia-Naya,~\IEEEmembership{Member,~IEEE},
	and~Luis Castedo,~\IEEEmembership{Senior Member,~IEEE} %
	\thanks{T.~ Dom\'inguez-Bola\~no, J. A. Garc\'ia-Naya, and L. Castedo are with the Universidade da Coruña (University of A Coru\~na), CITIC Research Center, A Coru\~na, Spain. (e-mail: tomas.bolano@udc.es, jagarcia@udc.es, and luis@udc.es).}
	\thanks{J.~Rodr\'iguez-Pi\~neiro is with the College of Electronics and Information Engineering, Tongji University, Shanghai, China. (e-mail: j.rpineiro@tongji.edu.cn).}
	\thanks{Corresponding author: Jos\'e~Rodr\'iguez-Pi\~neiro (e-mail address: j.rpineiro@tongji.edu.cn).}
	\thanks{
		This work has been funded by the Xunta de Galicia (ED431G2019/01), the Agencia Estatal de Investigación of Spain (TEC2016-75067-C4-1-R, PID2019-104958RB-C42), ERDF funds of the EU (AEI/FEDER, UE), and the
		National Natural Science Foundation of China (NSFC) under Grants no. 61850410529 and 61971313.}
	\thanks{Manuscript received XXX, XX, 2020; revised XXX, XX, 2020.}
	\thanks{Digital Object Identifier 10.1109/TVT.2020.3011365}
}

\markboth{IEEE Transactions on Vehicular Technology,~Vol.~XX, No.~XX, XXX~2020}
{}

\maketitle

\begin{abstract}
	Doubly-selective channels, such as those that occur when the transmitter and the receiver move relative to each other at high speeds, are a key scenario for \ac{5G} cellular systems, which are mostly based in the use of the \ac{OFDM} modulation. In this paper, we consider an \ac{OFDM} system using \ac{QAM} symbols and we show that, when transmitting over deterministic doubly-selective channels, the \ac{ICI} affecting a symbol can be well approximated by a complex-valued normal distribution. Based on this, we derive a lower bound for the link capacity using the Shannon-Hartley theorem. Finally, we provide an approximation of the \ac{BEP} using the well-known \ac{BEP} expressions for Gray-coded \ac{QAM} constellations over \ac{AWGN} channels, and show numerical results that confirm that the proposed \ac{BEP} expression approximates accurately the \ac{BER} of the \ac{OFDM} system for standardized channel models. The proposed closed-form analytical expressions for the capacity and the \ac{BEP} do not only allow for discarding the need of computationally-costly Monte-Carlo system simulations, but also provide a theoretical framework to optimize the system parameters directly impacting on the achievable performance.
\end{abstract}


\IEEEpeerreviewmaketitle
\acresetall

\positiontextbox{11cm}{27cm}{\footnotesize \textcopyright 2020 IEEE. This version of the article has been accepted for publication, after peer review. Personal use of this material is permitted. Permission from IEEE must be obtained for all other uses, in any current or future media, including reprinting/republishing this material for advertising or promotional purposes, creating new collective works, for resale or redistribution to servers or lists, or reuse of any copyrighted component of this work in other works. Published version:
\url{https://doi.org/10.1109/TVT.2020.3011365}}

\section{Introduction}

\IEEEPARstart{S}{ome} of the most successful standards for high-speed wireless data communications, such as Wi-Fi, \ac{LTE}, or the \ac{5G} cellular systems, among many others, rely on \ac{OFDM} to transmit information over the air. This widespread adoption of \ac{OFDM} was prompted by its large number of advantages such as its robustness against multipath propagation and its relatively low complexity.

In \ac{OFDM} the information is transmitted in parallel over a group of orthogonal narrowband subcarriers, employing conventional single-carrier modulation techniques such as \ac{QAM} or \ac{PSK}. The subcarriers used in \ac{OFDM} are just harmonic complex-valued exponentials, hence \ac{OFDM} systems are implemented in practice very efficiently by using the \ac{IFFT} for the modulator and the \ac{FFT} for the demodulator \cite{cho2010mimo}. \ac{OFDM} systems are able to maintain the orthogonality of the subcarriers in time-invariant multipath channels, i.e., in frequency-selective channels, very easily by using a \ac{CP} \cite{cho2010mimo}. In these cases, assuming that the maximum delay of the channel impulse response does not exceed the \ac{CP} length, the effect of the channel for each subcarrier comes down to a multiplication by a complex number, namely the coefficient of the channel impulse response in the frequency domain over that subcarrier.

However, in time-varying channels, the multipath components of the channel vary over time, thus destroying the orthogonality of the \ac{OFDM} subcarriers and introducing \ac{ICI} in the received signal. Hence, in these channels the performance of \ac{OFDM} systems can degrade significantly. When a channel is both time- and frequency-selective it is usually referred to as a doubly-selective channel.
Such channels occur in scenarios where the receiver and the transmitter move relative to each other, for example in high-speed train communications\cite{ai2014challenges,ArtigoMedidasTren_WCMC2017,zhang2014measurement,liu2012position,ArtigoPerdasPropagacion_Measurement2016}, \ac{V2V} and \ac{V2I} communications\cite{paier2009characterization,molisch2009survey,matolak2014modeling,mecklenbrauker2011vehicular,viriyasitavat2015vehicular}, or underwater acoustic communications\cite{walree2013propagation}.

One of the requirements for \ac{5G} cellular systems, as defined by the \ac{ITU}, is to support several mobility classes\cite{IMT-2020}, ranging from stationary up to high speed vehicular with a maximum speed of 500 km/h. In this regard, two important scenarios are the high speed train communications\cite{hasegawa2018high} and the \ac{UAV} communications \cite{li2018uav}. The importance of such cases for \ac{5G} systems was acknowledged by the \ac{3GPP} which has recently approved several study and work items for both use cases \cite{3GPP_LTE_high_speed_enh2,3GPP_NR_HST,3GPP_EAV}. Moreover, we can expect that high speed scenarios such as the aforementioned ones, and also novel ones, will be more and more important for the future \ac{6G} communication networks \cite{dang2020should}.

When assessing the performance of communication systems, one of the most used metrics is the \ac{BEP}. To analyze the \ac{BEP} one must first consider a suitable channel model for the desired use case being considered. We can distinguish the following two classes of channel models \cite{molisch2005wireless}:
\begin{itemize}
	\item \emph{Stochastic channel models} which usually assume that the channel is \ac{WSS} or \ac{WSSUS}, and that the statistics of some of the system functions, such as the channel impulse response, are known.
	\item \emph{Deterministic channel models}\footnote{We call \emph{deterministic models} to the \emph{site-specific models} as defined in \cite{molisch2005wireless}.} where the \ac{CIR} is known. In practice, the \ac{CIR} may come from realizations of some statistical channel model (e.g., a tapped delay line model), from measurements, or by solving Maxwell's equations (or some approximation) for a given environment.
\end{itemize}

There exist previous studies which analyze the performance of \ac{OFDM} in terms of the \ac{BEP} for stochastic doubly-selective channels. In this regard, the performance of  \ac{OFDM} was analyzed under the assumption of Gaussian \ac{ICI} in \cite{russell1995interchannel,wan2000bit,chiavaccini2000error,nissel2017ofdm}\footnote{For simplicity, the analysis in \cite{russell1995interchannel} considered the \ac{SEP} instead of the \ac{BEP}.}. However, for stochastic doubly-selective channels the \ac{PDF} of the \ac{ICI} is not Gaussian, but a weighted Gaussian mixture \cite{nissel2017ofdm, wang2006performance}. In this case, a Gaussian assumption allows for obtaining an approximation of the actual \ac{OFDM} performance, but it can actually lead to relatively large \ac{BEP} estimation errors \cite{nissel2017ofdm}. In \cite{wang2006performance}, the authors approximated the actual \ac{PDF} of the \ac{ICI} for a stochastic doubly-selective channel by means of a truncated Gram-Charlier series representation, which allowed them to obtain a more accurate expression (albeit much more complex) for the \ac{BEP}. Nevertheless, all these previous works only consider Rayleigh channel models.

\emph{Stochastic channel models} are usually preferred for design and comparison of wireless systems, whereas \emph{deterministic channel models} are more suitable for network planing and system deployment, where we must consider the site specific channel impulse responses \cite{molisch2005wireless}. In this regard, \emph{deterministic channel models} are used in link\cite{pratschner2018versatile,ArtigoSimuladorGTEC_IWSLS2016,nikaein2014openairinterface} and system level\cite{rupp2016vienna,ns3} simulations to determine performance metrics such as the \ac{BER} or the throughput for specific deployments of wireless systems. Recall that the \ac{BER} is calculated as the number of erroneous bits divided by the number of total bits transmitted. Hence, for a sufficiently large number of transmitted bits, the \ac{BER} becomes an approximation of the \ac{BEP}. For large complex systems, the \ac{BER} can be obtained by means of Monte-Carlo simulations, but this is a very computing-intensive process. Significant computational and time savings can be obtained if analytical expressions are used to determine exact or close approximations of the \ac{BEP}.

An approach considered by some of the previous works to calculate the \ac{BEP} over stochastic channels is to average the \ac{BEP} over several channel realizations, assuming that for each single channel realization the \ac{ICI} over each symbol is Gaussian when the number of subcarriers is large due to the \ac{CLT} \cite{russell1995interchannel,wan2000bit,chiavaccini2000error}. These works showed that the Gaussian assumption for individual channel realizations can yield accurate results of the \ac{BEP} when averaging over Rayleigh channels. However, if we want to employ this approach for deterministic channels, i.e., when not considering averaging over several channel realizations, several questions and problems arise that need to be investigated. Thus, this paper present several contributions which can be summarized in the following points:
\begin{enumerate}
	\item We show in \cref{sec:distribution_of_ICI} that, even if we consider a large number of subcarriers, the conditions to invoke the \ac{CLT} are actually not fulfilled, and the \ac{ICI} is not Gaussian.
	\item Next, we show in \cref{sec:distribution_of_ICI} that a Gaussian distribution may still be a good approximation for the \ac{ICI}, but the goodness of such approximation will depend on the constellation order of the transmit symbols (approximation improves with larger orders) as well as the subcarrier position (approximation gets worse on the border subcarriers).
	\item Based on the Gaussian assumption of the \ac{ICI}, an approximation for the \ac{BEP} for deterministic channels as well as a bound on the capacity can be easily obtained from the well-known formulas for the \ac{AWGN} channel. We show this in \cref{sec:capacity_and_bep}.
	\item The Gaussian assumption may lead to good results when averaging the \ac{BEP} for Rayleigh channels. However, since the \ac{ICI} is not exactly Gaussian it is not clear if good approximations can be also obtained for the case of deterministic channels, e.g., single channel realizations. Thus, we devote \cref{sec:results_A} to characterize the goodness of the \ac{BEP} approximation for deterministic channels by means of the error quotient $\rho = \mathrm{BER}/\mathrm{BEP}$, where the \ac{BER} is obtained by means of Monte-Carlo simulations. We show that for low \ac{SINR} values we obtain good approximations (the error quotient is close to 1), whereas for larger \acp{SINR} values, the approximation becomes worse but is always an upper bound of the \ac{BER} (the error quotient is less than 1).
	\item Finally, we calculate in \cref{sec:results_B} the average \ac{BEP} for an stochastic channel by averaging the \ac{BER} over several channel realizations, in a similar way as in previous works \cite{russell1995interchannel,wan2000bit,chiavaccini2000error}. Different from those works, we take into account that the Gaussian approximation of the \ac{ICI} depends on the constellation order and the subcarrier considered, and show results for different constellation orders and two different subcarriers, one central subcarrier where the Gaussian approximation of the ICI is better, and one border subcarrier where the Gaussian approximation of the \ac{ICI} is the worst. A novel results is that even considering a border subcarrier, the mean results of the \ac{BER} and the \ac{BEP} match for all the constellation orders with the exception of $4$-QAM for very low maximum Doppler shift values. Moreover, based on the characterization of the goodness of the \ac{BEP} approximation of \cref{sec:results_A} we are able to explain why averaging the \ac{BER} over several channel realizations yields good results.
\end{enumerate}


We also provide the source code employed to obtain the results included in this paper, which supports the reproducibility of the presented results and allows other researchers to easily apply our findings to their works.

\section{System Model} \label{sec:system_model}

Let us consider the transmission of \ac{OFDM} symbols over a point-to-point single-antenna wireless link. The length of the \ac{OFDM} symbols is $T+T_{\mathrm{cp}}$, where $T$ is the inverse of the subcarrier spacing, and $T_{\mathrm{cp}}$ is the length of the \ac{CP}. The $m$-th transmitted \ac{OFDM} symbol, starting at the time instant $t_m$ (not taking into account the \ac{CP}), can be expressed as
\begin{equation}
	u_m(t) = \sum_{l \in \mathcal{S}} X_{m,l} \mathrm{e}^{j2\pi l (t-t_m)/T }, \quad t_m \leq t < t_m + T
\end{equation}
where $\mathcal{S}$ is the set of used subcarrier indices, and $X_{m,l}$ is the symbol transmitted in the subcarrier $l$ and \ac{OFDM} symbol $m$. It is important for the reader to note here the difference between a symbol, i.e., $X_{m,l}$, and an \ac{OFDM} symbol, i.e., $u_m(t)$. We employ this nomenclature along the rest of the paper. 

We assume a doubly-selective tapped delay line channel model with a finite number of paths expressed as
\begin{equation} \label{eq:system_model}
	h(t,\tau) = \sum_{i=1}^{N} \alpha_i \mathrm{e}^{j2\pi \nu_i t} \delta(\tau - \tau_i)
\end{equation}
where $t$ and $\tau$ are the time and delay independent variables, respectively, $j$ is the imaginary unit, $\delta(\cdot)$ is the Dirac delta function, $N$ is the number of paths, and $\alpha_i$, $\nu_i$, and $\tau_i$ are, respectively, the complex-valued amplitude, the Doppler shift, and the delay corresponding to the $i$-th path. We assume that the maximum delay is less than the \ac{CP} length, i.e., $\max_i(\tau_i) < T_{\mathrm{cp}}$. Under this assumption, there is no \ac{ISI} affecting the signal at the receiver, and we can ignore the \ac{CP} for the calculations. Thus, the $m$-th received \ac{OFDM} symbol can be expressed as
\begin{align} \label{eq:received_symbol_time}
	 & y_m(t) = h(t,\tau) * u_m(t) + \sqrt{\frac{N_0}{2}} n(t) \notag\\
	& = \sum_{i=1}^{N} \alpha_i \mathrm{e}^{j2\pi \nu_i t} \sum_{l \in \mathcal{S}} X_{m,l} \mathrm{e}^{j2\pi l (t-\tau_i-t_m)/T } + \sqrt{\frac{N_0}{2}} n(t)
\end{align}
where $t_m \leq t < t_m + T$, $N_0$ is the power spectral density of the noise, $n(t)$ is a standard complex-valued Gaussian random process, and $*$ is the time-varying convolution operator defined as
\begin{equation}
	h(t,\tau) * u_m(t) = \int_{-\infty}^{\infty} h(t,\tau)u(t-\tau) \mathrm{d}\tau.
\end{equation}

For the $l$-th subcarrier of the $m$-th \ac{OFDM} symbol, the received symbol, $Y_{m,l}$, is obtained as
\begin{align} \label{eq:received_symbol_freq}
	Y_{m,l} & = \frac{1}{T}\int_{0}^{T} y_m(t+t_m) \mathrm{e}^{-j2\pi lt/T} \mathrm{d}t \notag \\
	        & = X_{m,l} H_{m,l} + \mathrm{ICI}_{m,l} + \sqrt{\frac{N_0}{2}}N_{m,l}
\end{align}
where $H_{m,l}$ is the corresponding channel coefficient, $\mathrm{ICI}_{m,l}$ is the \ac{ICI} coefficient, and $N_{m,l}$ is a standard complex-valued Gaussian random variable. The channel coefficient, $H_{m,l}$, can be expressed as
\begin{equation} \label{eq:H_coef}
	H_{m,l} = \sum_{i=1}^{N} H_{m,l}(\bm{\theta}_i)
\end{equation}
where $\bm{\theta}_i = [\tau_i, \nu_i, \alpha_i]$ is a vector with the $i$-th path parameters, and $H_{m,l}(\bm{\theta}_i)$ is the $i$-th path contribution to the channel coefficient, defined as
\begin{equation}
	H_{m,l}(\bm{\theta}_i) = \alpha_i \mathrm{e}^{-j2\pi l\tau_i/T} \mathrm{e}^{j2\pi \nu_i t_m} D(\nu_i)
\end{equation}
where the function $D(\cdot)$ is
\begin{equation} \label{eq:D}
	D(f) = \frac{1}{T}\int_{0}^{T}\mathrm{e}^{j2\pi ft}\mathrm{d}t = \begin{cases}
		1 & \text{if} \; f = 0 \\
		\dfrac{\mathrm{e}^{j2\pi fT} - 1}{j2\pi fT} & \text{if} \; f \neq 0.
	\end{cases}
\end{equation}

The \ac{ICI} term in \eqref{eq:received_symbol_freq} is defined as
\begin{equation}\label{eq:ICI}
	\mathrm{ICI}_{m,l} = \sum_{\substack{k \in \mathcal{S}\\ k \neq l}} X_{m,k} H_{m,l,k}^\mathrm{ICI}
\end{equation}
where $H_{m,l,k}^\mathrm{ICI}$ is the \ac{ICI} contribution of the $k$-th subcarrier to the $l$-th subcarrier during the $m$-th \ac{OFDM} symbol. In a similar way as in \cref{eq:H_coef}, the \ac{ICI} contribution of the $k$-th subcarrier to the $l$-th subcarrier during the $m$-th \ac{OFDM} symbol is
\begin{equation}
	H_{m,l,k}^\mathrm{ICI} = \sum_{i=1}^{N} H_{m,l,k}^\mathrm{ICI}(\bm{\theta}_i)
\end{equation}
where $H_{m,l,k}^\mathrm{ICI}(\bm{\theta}_i)$ is the \ac{ICI} contribution corresponding to the $i$-th path, defined as
\begin{equation} \label{eq:ici_terms}
	H_{m,l,k}^\mathrm{ICI}(\bm{\theta}_i) = \alpha_i \mathrm{e}^{-j2\pi k\tau_i/T} \mathrm{e}^{j2\pi \nu_i t_m} D\left(\tfrac{k-l}{T} + \nu_i\right).
\end{equation}


\section{Normal Approximation of the \ac{ICI} Distribution}
\label{sec:distribution_of_ICI}

We assume that the transmit symbols $X_{m,l}$ defined in the previous section are independent and identically distributed random variables which take values from a square \ac{QAM} constellation of order $M$ and have a variance of $\sigma_x^2$. Recall that the \ac{ICI} term for the $l$-th subcarrier and the $m$-th \ac{OFDM} symbol, $\mathrm{ICI}_{m,l}$ (see \cref{eq:ICI}), is expressed as a sum of products involving the transmit symbols ($X_{m,k}$) and the \ac{ICI} contributions between the different subcarriers ($H_{m,l,k}^\mathrm{ICI}$). Since we are considering deterministic channels, the coefficients $H_{m,l,k}^\mathrm{ICI}$ are known values for a given channel realization while the transmit symbols $X_{m,k}$ are random variables. Therefore, $\mathrm{ICI}_{m,l}$ is a weighted sum of random variables and, according to the \ac{CLT}, $\mathrm{ICI}_{m,l}$ may be approximated by a complex-valued normal distribution \cite{russell1995interchannel}. Note, however, that the terms of $\mathrm{ICI}_{m,l}$ are independent but not identically distributed. Indeed, the variance for each term in $\mathrm{ICI}_{m,l}$ is
\begin{equation}\label{eq:variance}
	\mathrm{Var}\left(X_{m,k} H_{m,l,k}^\mathrm{ICI}\right) = \sigma_x^2\left|H_{m,l,k}^\mathrm{ICI}\right|^2
\end{equation}
which, from \cref{eq:ici_terms} and \cref{eq:D}, decays with a factor proportional to the square of the subcarrier distance, i.e., $|k-l|^2$. Thus, only a few terms will contribute significantly to $\mathrm{ICI}_{m,l}$. Because of that, the assumptions of the \ac{CLT}, or any of its variants, will not be fully satisfied, and hence the distribution of $\mathrm{ICI}_{m,l}$ will not necessarily converge to normal as $|\mathcal{S}| \to  \infty$. Nevertheless, in the following we will experimentally show that a complex-valued normal distribution may be a good approximation to $\mathrm{ICI}_{m,l}$. For this matter, we will use the Mardia's skewness and kurtosis measures to test for multivariate normality\cite{mardia1970measures}.

For a random sample of size $n$ from a $p$-variate population, i.e., $\bm{Z}_i = \left({Z}_{1,i}, \dots, {Z}_{p,i}\right)$, $i=1,2,\dots,n$, Mardia defined the skewness and kurtosis, respectively, as
\begin{equation}\label{eq:b1p}
	b_{1,p} = \frac{1}{n^2}\sum_{i=1}^n \sum_{j=1}^n \left[ \left(\bm{Z}_i - {\bm{\bar{Z}}}\right)^T \bm{S}^{-1} \left(\bm{Z}_j - {\bm{\bar{Z}}}\right) \right]^3
\end{equation}
and
\begin{equation}\label{eq:b2p}
	b_{2,p} = \frac{1}{n}\sum_{i=1}^n \left[ \left(\bm{Z}_i - {\bm{\bar{Z}}}\right)^T \bm{S}^{-1} \left(\bm{Z}_i - {\bm{\bar{Z}}} \right) \right]^2
\end{equation}
where $\bm{S}$ is the sample covariance matrix and $\bm{\bar{Z}}$ the sample mean vector. 
From \cref{eq:b1p} and \cref{eq:b2p}, Mardia defined two statistics to test for normality. In our case, however, $\mathrm{ICI}_{m,l}$ does not follow exactly a normal distribution and hence the Mardia's test fails for large values of $n$. Thus, instead of testing for normality, we will check that the skewness and kurtosis values of $\mathrm{ICI}_{m,l}$ are relatively close to the expected ones for a normal distribution under realistic circumstances. In our case we have $p = 2$ with  $\bm{Z}_i = [ \Re\{\mathrm{ICI}_{m,l}\}, \Im\{\mathrm{ICI}_{m,l}\}]$. Then, according to \cite{mardia1970measures}, the expected values of skewness and kurtosis for a normal distribution are $\mathrm{E}[b_{1,2}] = 24/n$, and $\mathrm{E}[b_{2,2}] = 8(n-1)/(n+1)$, respectively. Thus, for large $n$ values, $\mathrm{E}[b_{1,2}] \approx 0$ and $\mathrm{E}[b_{2,2}] \approx 8$.

For our experiments we consider an \ac{OFDM} system with parameters similar to those of the \ac{LTE} $10$\,MHz downlink profile: the subcarrier spacing is $1/T = 15$\,kHz; the \ac{CP} length is $72/(15.36 \cdot 10^6)$ seconds (different \ac{CP} lengths are not considered for simplicity); and $600$ subcarriers are used, i.e., $\mathcal{S} = \{-300, \dots, -1, 1, \dots, 300\}$. We will consider $4$-QAM, $16$-QAM, and $64$-QAM constellations, and the following three channel models:
\begin{itemize}
	\item {3GPP typical urban channel model (TUx)}\cite[table 5.2]{3GPPTR25.943}: 20-taps channel model, presenting a high frequency selectivity and using the Jakes Doppler spectrum for all the taps.
	\item {3GPP rural area channel model (RAx)}\cite[table 5.3]{3GPPTR25.943}: 10-taps channel model, with a direct path, using the Jakes Doppler spectrum for the remaining taps.
	\item {ITU-R vehicular -- high antenna channel model}\cite[table 5]{ITURM.1225}: 6-tap channel model, using the Jakes Doppler spectrum for all the taps.
\end{itemize}

\begin{figure}[t]
	\centering
	\includegraphics[width=1\columnwidth]{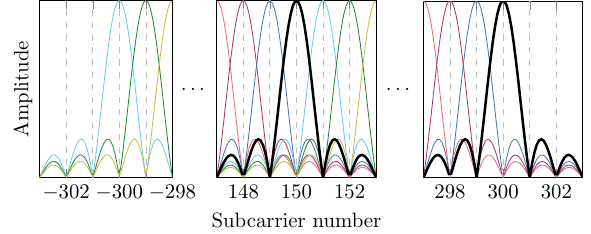}
	\caption{Frequency spectra of the subcarriers used for the evaluations. The spectra of subcarriers $l = 150$ and $l = 300$ correspond to the thick black curves.}
	\label{fig:Subcarriers}
\end{figure}

We generate channel realizations by using the Clarke's model with eight sinusoids for the taps which follow the Jakes Doppler spectrum\cite{xiao2006novel}, and a single sinusoid for the direct paths. Random samples of $\mathrm{ICI}_{m,l}$ are obtained by giving values to the symbols $X_{m,k}$ in \cref{eq:ICI}. For the sake of brevity, we present the results for two exemplary cases: \textit{a)} a middle subcarrier ($l = 150$), and \textit{b)} a border subcarrier ($l = 300$), as imaged in \cref{fig:Subcarriers}. Note that the so-called middle subcarrier is surrounded by other subcarriers used for the transmission of symbols, whereas the border subcarrier only has busy subcarriers at one side. This way, the middle subcarrier is expected to be the best-case scenario since the decay of the variance of the \ac{ICI} terms corresponding to the subcarriers at both sides of the middle-subcarrier is symmetric, according to \cref{eq:variance}. On the contrary, for the border subcarrier, only the subcarriers at one of its sides contribute to the \ac{ICI}, leading to a non-symmetrical distribution of the corresponding \ac{ICI} terms.

For a given channel realization and constellation order we generate $n = 10^3$ random samples of $\mathrm{ICI}_{0,150}$ and $\mathrm{ICI}_{0,300}$, and calculate the Mardia's skewness and kurtosis. Note that for $n = 10^3$ the expected values of skewness and kurtosis for normally distributed samples become $\mathrm{E}[b_{1,2}] = 0.024$ and $\mathrm{E}[b_{2,2}] = 7.984$, respectively. The previous process is performed for $10^3$ channel realizations for each of the three considered channel models and for several maximum normalized Doppler shifts\footnote{The maximum Doppler shift of a communication link is expressed as $\nu_{\mathrm{max}} = s f_c/c_0$, being $s$ the relative speed between the transmitter and the receiver, $f_c$ the carrier frequency, and $c_0$ the speed of light in the vacuum.} $\nu_{\mathrm{max}}T$. Note that for each channel realization we obtain a single value of skewness and kurtosis. Thus, with the $10^3$ considered channel realizations we have a set of $10^3$ values of skewness and kurtosis to be analyzed statistically.

\begin{table}[t]
	\caption{Skewness for $10^3$ random samples of $\mathrm{ICI}_{0,150}$ and $\mathrm{ICI}_{0,300}$ over $10^3$ channel realizations for $\nu_\mathrm{max}T = 0.05$}.
	\label{tab:ICI_skewness}
	\centering
	\begin{tabular}{crrrrr}
		\toprule
		Channel model & M & \multicolumn{2}{c}{$\mathrm{ICI}_{0,150}$ skewness} & \multicolumn{2}{c}{$\mathrm{ICI}_{0,300}$ skewness}\\
		\cmidrule(lr){3-4} \cmidrule(lr){5-6}
		& & mean & variance & mean & variance\\
		\midrule
		\multirow{3}{*}{3GPP TUx} & 4 & 0.0174 & 0.0002 & 0.0131 & 0.0001\\
		& 16 & 0.0187 & 0.0002 & 0.0154 & 0.0001\\
		& 64 & 0.0193 & 0.0002 & 0.0160 & 0.0001\\[0.5em]
		\multirow{3}{*}{3GPP RAx} & 4 & 0.0164 & 0.0001 & 0.0121 & 0.0001\\
		& 16 & 0.0182 & 0.0002 & 0.0140 & 0.0001\\
		& 64 & 0.0196 & 0.0002 & 0.0152 & 0.0001\\[0.5em]
		\multirow{3}{*}{ITU-R vehicular} & 4 & 0.0165 & 0.0002 & 0.0131 & 0.0001\\
		& 16 & 0.0182 & 0.0002 & 0.0149 & 0.0001\\
		& 64 & 0.0182 & 0.0002 & 0.0155 & 0.0001\\
		\bottomrule
	\end{tabular}
\end{table}

Mean and variance results for the skewness are shown in \cref{tab:ICI_skewness} considering a maximum normalized Doppler shift of $\nu_{\mathrm{max}}T = 0.05$, both for $\mathrm{ICI}_{0,150}$ and $\mathrm{ICI}_{0,300}$. The obtained results for the skewness are close to $0$ in all the cases, ranging from $0.01$ to $0.02$, hence indicating a high degree of symmetry of the \ac{ICI} distribution. Mean and variance results for the kurtosis are shown in \cref{tab:ICI_kurtosis}, again considering a maximum normalized Doppler shift of $\nu_{\mathrm{max}}T = 0.05$ for both $\mathrm{ICI}_{0,150}$ and $\mathrm{ICI}_{0,300}$. In this case the kurtosis values are similar for the three channel models, but there are differences with respect to the constellation order and subcarrier considered. Regarding the constellation order, the kurtosis values increase as the constellation order increases, being the biggest step the one from 4- to 16-QAM. With respect to the subcarrier considered, for $\mathrm{ICI}_{0,150}$ the obtained mean kurtosis ranges from $7.19$ to $7.54$ (approx.), whereas for $\mathrm{ICI}_{0,300}$ the obtained values are significantly lower, ranging from $6.35$ to $7.07$ (approx.). In all cases the mean kurtosis is less than $8$, stating that the \ac{ICI} does not exactly follow a normal distribution. However, values close to $8$ indicate that a normal distribution may still be a good approximation. For instance, consider the Student's $t$-distribution, which for $30$ degrees of freedom is a very good approximation to a normal distribution. A 2-dimensional multivariate Student's $t$-distribution with $30$ degrees of freedom has a kurtosis value of approximately $8.46$. Therefore, we may expect to have a good normal approximation of the \ac{ICI} for middle subcarriers, whereas for border subcarriers the approximation becomes rougher.

\begin{table}[t]
	\caption{Kurtosis for $10^3$ random samples of $\mathrm{ICI}_{0,150}$ and $\mathrm{ICI}_{0,300}$ over $10^3$ channel realizations for $\nu_\mathrm{max}T = 0.05$}.
	\label{tab:ICI_kurtosis}
	\centering
	\begin{tabular}{crrrrr}
		\toprule
		Channel model & M & \multicolumn{2}{c}{$\mathrm{ICI}_{0,150}$ kurtosis} & \multicolumn{2}{c}{$\mathrm{ICI}_{0,300}$ kurtosis}\\
		\cmidrule(lr){3-4} \cmidrule(lr){5-6}
		& & mean & variance & mean & variance\\
		\midrule
		\multirow{3}{*}{3GPP TUx} & 4 & 7.2665 & 0.0608 & 6.5101 & 0.1882\\
		& 16 & 7.4881 & 0.0450 & 6.9844 & 0.0983\\
		& 64 & 7.5404 & 0.0510 & 7.0743 & 0.0866\\[0.5em]
		\multirow{3}{*}{3GPP RAx} & 4 & 7.1871 & 0.0275 & 6.3545 & 0.0181\\
		& 16 & 7.4474 & 0.0318 & 6.8759 & 0.0223\\
		& 64 & 7.4951 & 0.0378 & 6.9837 & 0.0265\\[0.5em]
		\multirow{3}{*}{ITU-R vehicular} & 4 & 7.2407 & 0.0500 & 6.5176 & 0.1654\\
		& 16 & 7.4714 & 0.0457 & 6.9840 & 0.0883\\
		& 64 & 7.5219 & 0.0439 & 7.0706 & 0.0841\\
		\bottomrule
	\end{tabular}
\end{table}

\begin{figure}[t]
	\begin{subfigure}[t]{0.47\columnwidth}
		\centering
		\newsavebox{\tempfig}
		\savebox{\tempfig}{\includegraphics[scale=1.1]{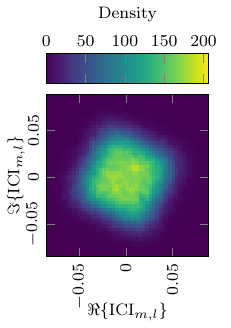}} 
		\raisebox{\dimexpr\ht\tempfig-\height}{\includegraphics[scale=1.1]{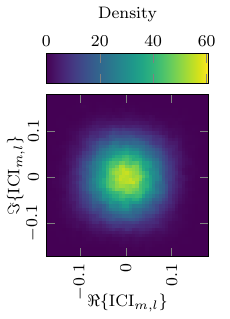}}
		\caption{Joint PDF of $\mathrm{ICI}_{0,150}$ with kurtosis = 7.243.}
		\label{fig:ICI_PDF_1_2}
	\end{subfigure}
	\hfill
	\begin{subfigure}[t]{0.47\columnwidth}
		\centering
		\includegraphics[scale=1.1]{fig02a.pdf}
		\caption{Joint PDF of $\mathrm{ICI}_{0,300}$ with kurtosis = 6.481.}
		\label{fig:ICI_PDF_1_1}
	\end{subfigure}
	\caption{Exemplary joint PDFs of the real and imaginary parts of $\mathrm{ICI}_{0,150}$ and $\mathrm{ICI}_{0,300}$ for 4-QAM.}
	\label{fig:ICI_PDF_1}
\end{figure}

In \cref{fig:ICI_PDF_1} we show exemplary joint \acp{PDF} of the \ac{ICI} for subcarriers 150 and 300 for a given channel realization, considering the ITU-R vehicular channel with $\nu_\mathrm{max}T = 0.05$ and a 4-QAM constellation. The kurtosis values for the joint \acp{PDF} shown in \cref{fig:ICI_PDF_1_1,fig:ICI_PDF_1_2} are $7.243$ and $6.481$, respectively. These values are similar to the mean ones for subcarriers $150$ and $300$ (see \cref{tab:ICI_kurtosis}). It can be seen that the \ac{PDF} of $\mathrm{ICI}_{0,300}$ has a squarish shape and thus is clearly non normal. On the other hand, $\mathrm{ICI}_{0,150}$ has a larger kurtosis value and the joint \ac{PDF} is much closer to a complex-valued normal distribution.

\begin{figure}[t]
	\centering
	\includegraphics[width=1\columnwidth]{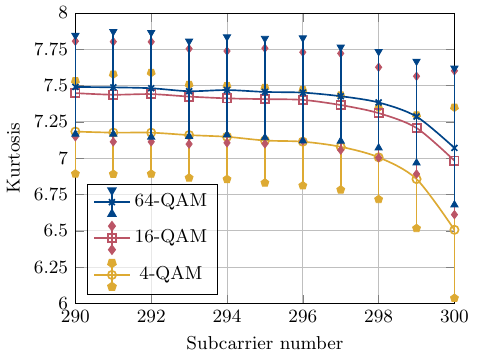}
	\caption{Means and $5\%$ and $95\%$ percentiles of the kurtosis values vs subcarrier number for $10^3$ realizations of the ITU-R vehicular channel model with $\nu_\mathrm{max}T = 0.05$.}
	\label{fig:ICI_vs_subcarrier}
\end{figure}

It its important to note that, when moving away from the border subcarriers, the kurtosis values increase very rapidly. This is because, as explained before, the variance of the \ac{ICI} decays with a factor proportional to the square of the subcarrier distance and only the closest subcarriers contribute significantly to the \ac{ICI}. In \cref{fig:ICI_vs_subcarrier} we show the means of the kurtosis values, as well as their $5\%$ and $95\%$ percentiles, obtained for different subcarriers considering an ITU-R channel model, $10^3$ channel realizations, and a normalized maximum Doppler shift of 0.05. For the sake of conciseness, we will show only results of the kurtosis for a single channel model. Results for the other channel models are similar. It can be seen that, starting at the border subcarrier, when the subcarrier number decreases by just three or four units, the mean kurtosis values are very close to the ones shown in \cref{tab:ICI_kurtosis} for subcarrier 150. Recall that, for each channel realization, the kurtosis value might be different, hence, the $5\%$ and $95\%$ percentiles are also plotted in \cref{fig:ICI_vs_subcarrier} to show the range of the $90\%$ central values of kurtosis obtained. It can be seen that for the $4$-QAM case the kurtosis values can reach sometimes much lower values than the ones expected for a normal distribution, specially on the border subcarriers, but for higher constellation orders the values are significantly larger.

\begin{figure}[t]
	\centering
	\includegraphics[width=1\columnwidth]{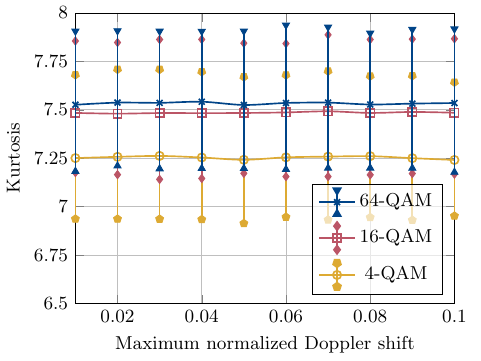}
	\caption{Means and $5\%$ and $95\%$ percentiles of the kurtosis values vs maximum normalized Doppler shift for $10^3$ realizations of the ITU-R vehicular channel model and subcarrier $150$.}
	\label{fig:ICI_vs_max_Doppler}
\end{figure}

Finally, we investigate the effect of the maximum Doppler shift (i.e., the relative speed between the transmitter and the receiver) on the Mardia's skewness and kurtosis values. We find that such values do not present significative differences for any of the cases (i.e., channel model, constellation order, and subcarrier number) when the maximum Doppler shift varies. For the sake of conciseness, we only show results of the kurtosis for a single channel model and subcarrier. \cref{fig:ICI_vs_max_Doppler} plots the means of the kurtosis values, as well as their $5\%$ and $95\%$ percentiles, obtained for different values of the maximum normalized Doppler shift, ranging from $0.01$ to $0.1$. The values are calculated for the subcarrier $150$ using $10^3$ channel realizations of the ITU-R vehicular channel model for each value of maximum normalized Doppler shift. It can be seen that neither the means nor the percentile values experience significant variations for any of the considered maximum normalized Doppler shift values.

\section{Capacity Bound and Bit Error Probability}
\label{sec:capacity_and_bep}

In the previous section we have shown that, although $\mathrm{ICI}_{m,l}$ does not follow a normal distribution, the normal distribution is still a good approximation to model the \ac{ICI} in practical cases. More specifically, we have seen that the approximation is only noticeably worse for the subcarriers near the edges of the effectively used bandwidth. This way, assuming that the \ac{ICI} distribution is well-approximated by a complex-valued normal distribution, the \ac{ICI} plus the noise for a given symbol can be also approximated by a complex-valued normal distribution. The variance of the distribution of the \ac{ICI} plus the noise for the symbol at time $m$ and subcarrier $l$ is
\begin{equation}\label{eq:ICI_N0}
	\sigma^2_{\mathrm{ICI}_{m,l}} = \mathrm{Var}(\mathrm{ICI}_{m,l}) + N_0
\end{equation}
where, from \cref{eq:ICI} and considering that the transmit symbols are mutually independent, the variance of $\mathrm{ICI}_{m,l}$ is
\begin{equation}\label{eq:Var_ICI}
	\mathrm{Var}(\mathrm{ICI}_{m,l}) =  \sigma_x^2 \sum_{\substack{k \in \mathcal{S}\\ k \neq l}} \left| H_{m,l,k}^\mathrm{ICI} \right|^2.
\end{equation}
The transmitted energy per bit, considering a square \ac{QAM} constellation of order $M$, is defined as
\begin{equation}
	E_b^{\mathrm{TX}} = \sigma_x^2/\log_2{M}.
\end{equation}
Based on this definition, the received energy per bit for the symbol at time $m$ and subcarrier $l$ is defined as
\begin{equation}
	E^{\mathrm{RX}}_{b_{m,l}} = E_b^{\mathrm{TX}}\left| H_{m,l} \right|^2.
\end{equation}
Then, the ratio of the received energy per bit to the noise plus interference power spectral density is
\begin{equation}\label{eq:ratio_Eb_ICINO}
	r_{m,l} = \frac{E^{\mathrm{RX}}_{b_{m,l}}}{\sigma^2_{\mathrm{ICI}_{m,l}} } = \frac{\sigma_x^2 \left| H_{m,l} \right|^2}{\sigma^2_{\mathrm{ICI}_{m,l}} \log_2{M}}.
\end{equation}

The capacity\footnote{We define the capacity as the maximum rate at which we can transmit information over a given channel  with our system  (i.e., the achievable throughput per Hertz).} can be bounded from \cref{eq:ratio_Eb_ICINO} by using the Shannon-Hartley theorem. This is because the capacity of a channel with additive non-Gaussian noise is greater than or equal to the capacity of a channel with additive Gaussian noise, assuming that the noise covariances are the same \cite{IHARA197834}. Therefore, the capacity for the symbol at time $m$ and subcarrier $l$, with units of bits/s/Hz, can be expressed as the following lower bound
\begin{equation}
	\label{eq:capacity}
	C_{m,l} \geq \frac{T}{T + T_{\mathrm{cp}}} \log_2\left(1 + r_{m,l}\log_2{M}\right)
\end{equation}
where the factor $T/(T+T_{\mathrm{cp}})$ accounts for the spectral efficiency loss due to the \ac{CP}, and $r_{m,l}\log_2{M}$ is the ratio of the received  energy  per symbol to  the noise plus interference power spectral density.


In the same way, by using \cref{eq:ratio_Eb_ICINO} and considering the normal approximation, the \ac{BEP} for a given symbol can be approximated analytically. First, note that the exact closed-form \ac{BEP} expression for a communication system employing an $M$-ary square \ac{QAM} constellation with Gray mapping over an \ac{AWGN} channel is \cite{883298}
\begin{equation}
	P_M(r) = \frac{1}{\log_2\sqrt{M}}\sum_{j=1}^{\log_2\sqrt{M}} P_{M,j}(r)
\end{equation}
where $r$ is the ratio of the energy per bit to the noise power spectral density, and the $P_{M,j}(r)$ is the error probability of the $j$-th bit for the $M$-QAM constellation, defined as
\begin{align}
	P_{M,j}(r) &= \frac{1}{\sqrt{M}} \sum_{k=0}^{\substack{(1-2^{-j})\\\cdot \sqrt{M} - 1}}
	\Bigg[(-1)^{\big\lfloor {k \cdot 2^{j-1}}/{\sqrt{M}} \big\rfloor} \nonumber \\
	& \cdot\left( 2^{j-1} - \bigg\lfloor \frac{k \cdot 2^{j-1}}{\sqrt{M}} + \frac{1}{2} \bigg\rfloor \right) \nonumber \\
	& \cdot\mathrm{erfc}\Bigg( (2k + 1)\sqrt{\frac{3\log_2{M} \cdot r}{2(M - 1)}} \Bigg) \Bigg]
\end{align}
where $\lfloor \cdot \rfloor$ is the floor function and $\mathrm{erfc}(\cdot)$ is the complementary error function \cite{Andrews97special}. Hence, the analytical \ac{BEP} approximation when using an $M$-ary square QAM constellation for the symbol at time $m$ and subcarrier $l$ is simply obtained as
\begin{equation} \label{eq:BEP_approx}
	\mathrm{BEP}_{m,l} = P_M(r_{m,l}).
\end{equation}

However, note that, although the capacity expression in \cref{eq:capacity} is a lower bound, this does not imply that the analytical \ac{BEP} approximation in \cref{eq:BEP_approx} is an upper bound, because the receiver is optimal for Gaussian noise but now we have Gaussian noise plus non-Gaussian interference, which can make the receiver suboptimal \cite{lapidoth1996nearest}.

\section{Results}\label{sec:results}

In the following we show results of the analytical \ac{BEP} approximation as per \cref{eq:BEP_approx} and results of \ac{BER}
obtained by means of Monte-Carlo computer simulations. Note that we will not show experimental results of the capacity due to the complexity of the mathematical expressions and algorithms required to estimate the capacity for the non-Gaussian case.


For the results we consider $\sigma_x = 1$, and that the average channel gain is also $1$. For the simulations, several iterations are performed over each given channel realization to obtain averaged \ac{BER} results. Assuming that we want to calculate the \ac{BER} for a symbol at time $m$ and subcarrier $l$, an iteration of the simulation involves the following steps:
\begin{enumerate}
	\item Generate a sequence of random bits.
	\item Encode the bits to the $M$-QAM transmit symbols $X_{m,k}$ using Gray coding. Note that even though we only want to obtain the \ac{BER} for the subcarrier $l$, we must consider the transmission of symbols over all subcarriers.
	\item Apply the channel response and add the noise to the transmit symbols to obtain the receive symbol $Y_{m,l}$ as per \cref{eq:received_symbol_freq}.
	\item Obtain an estimate of the transmit symbol $X_{m,l}$. We consider perfect channel knowledge at the receiver and we employ a zero forcing estimation. Hence, the estimated receive symbol is obtained as $\hat{X}_{m,l} = Y_{m,l}/H_{m,l}$.
	\item Decode the estimated symbol $\hat{X}_{m,l}$ using Gray decoding to obtain the corresponding sequence of bits.
	\item Compare the received bits with the corresponding transmitted bits to obtain the \ac{BER} for the iteration.
\end{enumerate}

%

\begin{figure}[t]
	\centering
	\begin{subfigure}[t]{\linewidth}
		\includegraphics[width=0.96\columnwidth]{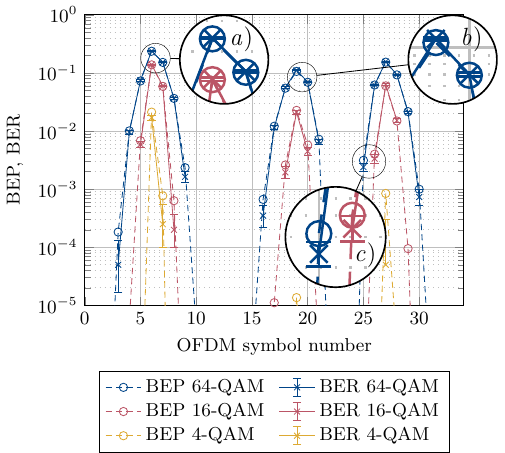}
		\caption{\ac{BER} (including the $95\%$ confidence intervals) and \ac{BEP} results.}
		\label{fig:bep_vs_symbol}
	\end{subfigure}
	\par\bigskip
	\begin{subfigure}[t]{\linewidth}
		\includegraphics[width=0.90\columnwidth]{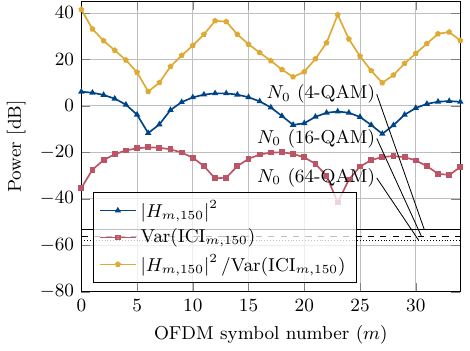}
		\caption{Channel power, \ac{ICI} power, and ratio of the channel power to the \ac{ICI} power.}
		\label{fig:H_and_ICI_powers}
	\end{subfigure}
	\caption{Exemplary results for subcarrier $150$, $35$ \ac{OFDM} symbols, and a single realization of the ITU-R vehicular channel model with $\nu_\mathrm{max}T = 0.05$.}
	\label{fig:Exemplanary_results}
\end{figure}

In \cref{fig:bep_vs_symbol} we show some exemplary results of the \ac{BEP} and \ac{BER} versus the \ac{OFDM} symbol number for a single realization of the ITU-R vehicular channel. We considered the subcarrier 150, 35 consecutive \ac{OFDM} symbols, a maximum normalized Doppler shift of $0.05$, and $10^4$ iterations for the \ac{BER} results.
The ratio of the energy per bit on transmission to the noise spectral power density, i.e., $E_b^{\mathrm{TX}}/N_0$, is set to $50$\,dB.
Note that $E_b^{\mathrm{TX}}/N_0$ is employed to obtain the noise value in the results, except those shown in \cref{fig:error_factors_subc_150,fig:error_factors_subc_300}, since the energy per bit at reception, as well as the power of the \ac{ICI}, depend on the \ac{OFDM} symbol, the subcarrier, and the channel realization.

We also show in \cref{fig:bep_vs_symbol} the $95\%$ \acp{CI} of the \ac{BER} results obtained by means of a bootstrapping technique \cite{bBootstrap_Efron2004}. It can be seen that the \ac{BEP} presents large variations over the time, due to the variations of the channel response and the \ac{ICI} powers. In \cref{fig:H_and_ICI_powers} we show the corresponding values of the channel and ICI power, as well as their ratio, versus the \ac{OFDM} symbol number. Note that since we are considering a constant symbol energy, $\sigma_x = 1$, the bit energy depends on the constellation order, and thus the noise power $N_0$ also depends on the constellation order, satisfying $E_b^{\mathrm{TX}}/N_0 = 50$\,dB. The corresponding values of the noise power $N_0$ for the different constellation orders are also shown in \cref{fig:H_and_ICI_powers}. It can be seen that the $E_b^{\mathrm{TX}}/N_0$ value of $50$\,dB is large enough so that the noise is significantly lower than the \ac{ICI} for all the \ac{OFDM} symbols. Therefore, the \ac{BEP} and \ac{BER} values shown in \cref{fig:bep_vs_symbol} are mostly due to the \ac{ICI} and not the noise.

Even though the results in \cref{fig:bep_vs_symbol} correspond to a single channel realization, they show already several interesting features that we comment below 
\begin{itemize}
	\item For some symbols, the ratio of the received energy per bit to the noise plus interference is very large, which means that for these symbols we obtained a \ac{BER} of zero. This is notorious in the $4$-QAM case.
	\item For the largest \ac{BER} values, the \ac{BEP} matches the \ac{BER} very well, e.g., see magnification \emph{a}) in \cref{fig:bep_vs_symbol}. In some cases the \ac{BER} may be slightly larger that the \ac{BEP}, e.g., see magnification \emph{b}) in \cref{fig:bep_vs_symbol}.
	\item For lower \ac{BER} values, the \ac{BEP} values are larger, and in some cases even larger than the upper $95\%$ \ac{CI}. Nevertheless, most of the \ac{BER} values are still close to the \ac{BEP} values, e.g., see magnification \emph{c}) in \cref{fig:bep_vs_symbol}.
\end{itemize}

\subsection{Instantaneous BEP and BER}
\label{sec:results_A}

Based on the previous remarks, it can be seen that the goodness of the \ac{BEP} approximation, i.e., how close is the \ac{BEP} as defined in \cref{eq:BEP_approx} to the \ac{BER}, may depend on the ratio of the received energy per bit to the noise plus interference.
To characterize this behavior, we define the approximation error factor metric for a given symbol as
\begin{equation}\label{ec:rho}
	\rho = \frac{\mathrm{BER}}{\mathrm{BEP}}.
\end{equation}

\begin{figure}[t]
	\centering
	\begin{subfigure}[b]{0.95\columnwidth}
		\includegraphics[width=\linewidth]{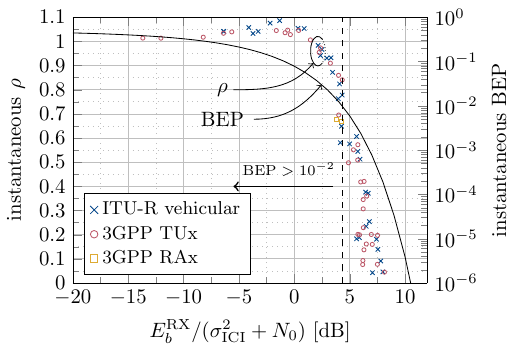}
		\caption{$4$-QAM.}
		\label{fig:error_factors_subc_150_a}
	\end{subfigure}
	\begin{subfigure}[b]{0.95\columnwidth}
		\includegraphics[width=\linewidth]{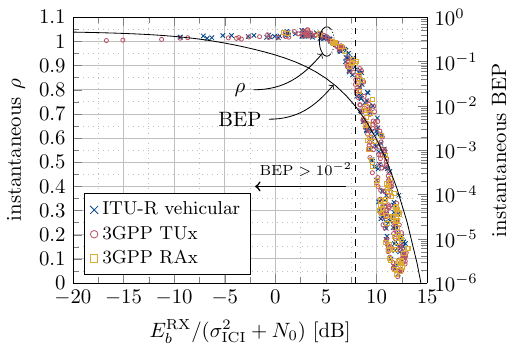}
		\caption{$16$-QAM.}
		\label{fig:error_factors_subc_150_b}
	\end{subfigure}
	\begin{subfigure}[b]{0.95\columnwidth}
		\includegraphics[width=\linewidth]{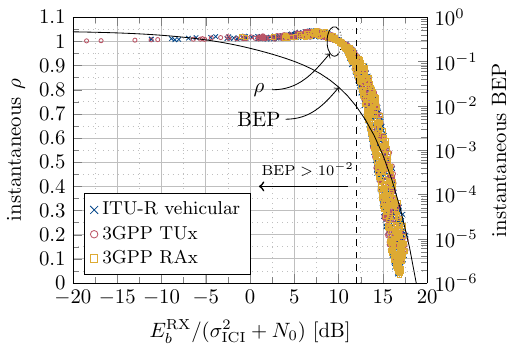}
		\caption{$64$-QAM.}
		\label{fig:error_factors_subc_150_c}
	\end{subfigure}
	\caption{Instantaneous error factors, $\rho = {\mathrm{BER}}/{\mathrm{BEP}}$, and instantaneous \ac{BEP} versus the ratio of the received energy per bit to the noise plus \ac{ICI} for subcarrier $150$, $\nu_\mathrm{max}T = 0.05$, and $100$ channel realizations for each of the considered channel models.}
	\label{fig:error_factors_subc_150}
\end{figure}

We calculate the error factors for the same three constellation orders and three channel models considered in \cref{sec:distribution_of_ICI}, for subcarrier 150, 10 consecutive \ac{OFDM} symbols, a maximum Doppler shift of 0.05, and $100$ different channel realizations. As in the results in \cref{fig:bep_vs_symbol}, we also consider $E_b^{\mathrm{TX}}/N_0$ = $50$\,dB. For each transmitted symbol the \ac{BEP} is obtained as per \cref{eq:BEP_approx}, and the \ac{BER} is estimated using $10^6$ iterations (each iteration is specified at the beginning of \cref{sec:results}). Moreover, to avoid possible outliers and to reduce the variance of the results shown, we discard the \ac{BER} estimates that lead to less than $10$ erroneous bits. Finally, the $\rho$ value corresponding to the \ac{BER} estimate is computed according to \cref{ec:rho}. Note that here we are considering the ``instantaneous'' results, i.e., the values are obtained for single symbols and no averaging between symbols is performed.

\begin{figure}[t]
	\centering
	\includegraphics[width=1\columnwidth]{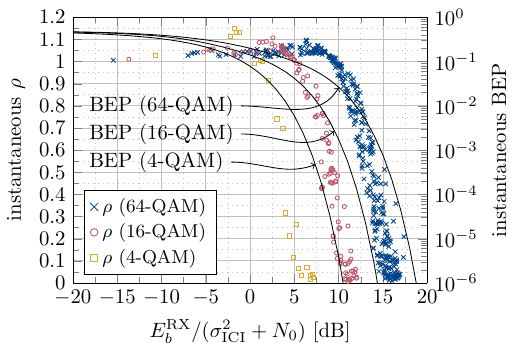}
	\caption{Instantaneous error factors, $\rho = {\mathrm{BER}}/{\mathrm{BEP}}$, and instantaneous \ac{BEP} versus the ratio of the received energy per bit to the noise plus \ac{ICI} for subcarrier $300$, $\nu_\mathrm{max}T = 0.05$, and $100$ channel realizations of the ITU-R vehicular channel.}
	\label{fig:error_factors_subc_300}
\end{figure}

The results obtained are shown in \cref{fig:error_factors_subc_150} as a cloud of points of the the error factors $\rho$ versus the ratio of the received energy per bit to the noise plus interference, i.e., $E^\mathrm{RX}_b/(\sigma^2_{\mathrm{ICI}} + N_0)$, for each of the symbols. Moreover, the \ac{BEP} curve as per \cref{eq:BEP_approx} is also shown for the considered range of $E^\mathrm{RX}_b/(\sigma^2_{\mathrm{ICI}} + N_0)$. Note that, as shown in \cref{fig:ICI_vs_max_Doppler}, the goodness of approximation of the \ac{ICI} by a Gaussian distribution is independent of the maximum normalized Doppler shift. Thus, the results shown in \cref{fig:error_factors_subc_150,fig:error_factors_subc_300} only depend on $E^\mathrm{RX}_b/(\sigma^2_{\mathrm{ICI}} + N_0)$. It can be seen that the results obtained agree with the aforementioned remarks made for the results in \cref{fig:bep_vs_symbol}. We can extract the following conclusions from the results shown in \cref{fig:error_factors_subc_150}:
\begin{itemize}
	\item There is no significant difference in the error factors obtained for the different channel models considered. This is the expected result since the goodness of the normal approximation of the \ac{ICI} also does not exhibit significant changes for the different channel models, as shown in \cref{sec:distribution_of_ICI}. Moreover, as shown in \cref{fig:ICI_vs_max_Doppler}, the goodness of the normal approximation of the \ac{ICI} does not depend on the maximum normalized Doppler shift. Thus, the results shown in \cref{fig:error_factors_subc_150} will also be the same for different values of the maximum normalized Doppler shift.
	\item For low $E^\mathrm{RX}_b/(\sigma^2_{\mathrm{ICI}} + N_0)$ values, $\rho$ is close to $1$. Thus, the \ac{BEP} results approximate well the \ac{BER}.
	\item It can be seen that for $16$-QAM and $64$-QAM, $1.1 < \rho < 0.8$ when $\mathrm{BEP} < 10^{-2}$. For $4$-QAM, due to the worse normal approximation of the \ac{ICI}, $1.1 < \rho < 0.5$ when $\mathrm{BEP} < 10^{-2}$.
	\item For higher $E^\mathrm{RX}_b/(\sigma^2_{\mathrm{ICI}} + N_0)$ values, i.e., when $\mathrm{BEP} > 10^{-2}$, $\rho$ is always less than one and monotonically decreases as $E^\mathrm{RX}_b/(\sigma^2_{\mathrm{ICI}} + N_0)$ increases. Thus, the \ac{BER} is always lower than the $\mathrm{BEP}$, i.e., the $\mathrm{BEP}$ becomes an upper bound of the \ac{BER}.
\end{itemize}

In \cref{fig:error_factors_subc_300} we show results of error factors for subcarrier $300$ and the ITU-R vehicular channel model. Results for the other channel models are similar and we do not show them for conciseness. We can see that these results are similar to the ones shown in \cref{fig:error_factors_subc_150} for subcarrier $150$. However, for all the constellation orders it can be seen that for low $E^\mathrm{RX}_b/(\sigma^2_{\mathrm{ICI}} + N_0)$ values the error factors obtained are slightly higher than those in \cref{fig:error_factors_subc_150}. Also, for the highest $E^\mathrm{RX}_b/(\sigma^2_{\mathrm{ICI}} + N_0)$ values the error factors are lower than for subcarrier $150$, which indicates that the \ac{BEP} approximation is worse in this case.

\begin{figure}[t]
	\centering
	\includegraphics[width=0.98\columnwidth]{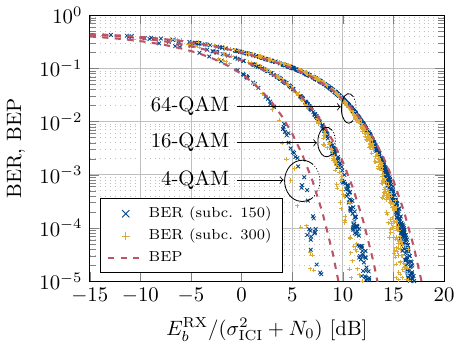}
	\caption{Instantaneous BER and BEP versus the ratio of the received energy per bit to the noise plus \ac{ICI} for subcarriers $150$ and $300$, $\nu_\mathrm{max}T = 0.05$, and $100$ channel realizations of the ITU-R vehicular channel.}
	\label{fig:ber_bep}
\end{figure}

In \cref{fig:ber_bep} we show the results of the instantaneous BER and BEP for subcarriers $150$ and $300$ and the ITU-R vehicular channel model. These results correspond to the ones shown in \cref{fig:error_factors_subc_150_a,fig:error_factors_subc_300}. \cref{fig:ber_bep} allows us to better appreciate the differences between the BEP and BER results obtained. We can see clearly that, as the $E^\mathrm{RX}_b/(\sigma^2_{\mathrm{ICI}} + N_0)$ increases, the \ac{BEP} becomes an upper bound of the \ac{BER}. We can also see that the \ac{BEP} approximation for the subcarrier $300$ is clearly worse than for the subcarrier $150$.

\begin{figure}[t]
	\centering
	\includegraphics[width=0.9\columnwidth]{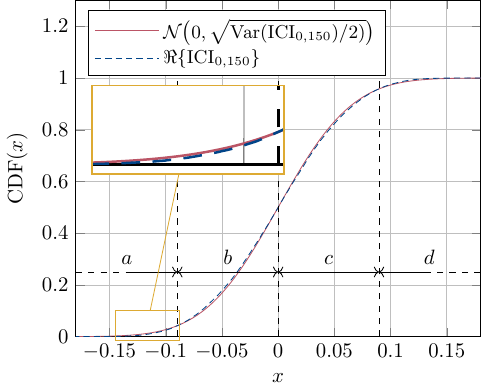}
	\caption{CDFs of the \ac{ICI} and the corresponding normal approximation for subcarrier 150 and a single channel realization of the ITU-R vehicular channel.}
	\label{fig:cdf_ber_bep_subc_150}
\end{figure}

The behavior of the \ac{BEP} approximation shown in the previous figures is due to the non-Gaussianity of the ICI distribution, more specifically because the ICI distribution is slightly platykurtic, i.e., its kurtosis is slightly lower than that of the Gaussian distribution. This can be clearly seen if we compare the \acp{CDF} of the \ac{ICI} and the Gaussian distributions. In \cref{fig:cdf_ber_bep_subc_150} we show the \ac{CDF} for the real part of the \ac{ICI} for subcarrier 150 and a single channel realization of the ITU-R vehicular channel. The channel realization corresponds to the one shown also in \cref{fig:ICI_PDF_1_2}. We also show the \ac{CDF} of the normal distribution which would be used to obtain the \ac{BEP} approximation. We define $\mathrm{CDF}_{\mathrm{ICI}}(\cdot)$ as the \ac{CDF} of the \ac{ICI} and $\mathrm{CDF}_{\mathcal{N}}(\cdot)$ as the \ac{CDF} of the normal distribution shown in \cref{fig:cdf_ber_bep_subc_150}. Due to the inferior kurtosis of the \ac{ICI} distribution, intervals $a$, $b$, $c$, and $d$ can be found such that:
\begin{align}
	\label{eq:cdf_ineq_1}
	\mathrm{CDF}_{\mathrm{ICI}}(x) \leq \mathrm{CDF}_{\mathcal{N}}(x) \quad \forall \, x \in a \, \vee  x \in c \\
	\label{eq:cdf_ineq_2}
	\mathrm{CDF}_{\mathrm{ICI}}(x) \geq \mathrm{CDF}_{\mathcal{N}}(x) \quad \forall \, x \in b \, \vee  x \in d
\end{align}
Thus, from \cref{eq:cdf_ineq_1,eq:cdf_ineq_2} we can clearly see that for low values of $E^\mathrm{RX}_b/(\sigma^2_{\mathrm{ICI}} + N_0)$ the \ac{BEP} approximation will be an upper bound, whereas for higher values of  $E^\mathrm{RX}_b/(\sigma^2_{\mathrm{ICI}} + N_0)$ the \ac{BEP} approximation will be slightly higher than the \ac{BER}, as shown in the previous results.

\subsection{Average BEP and BER}
\label{sec:results_B}

Some works \cite{russell1995interchannel,wan2000bit,chiavaccini2000error} have previously shown that a good approximation of the \ac{BEP} can be obtaining for stochastic channel models by averaging the \ac{BER} over several channel realizations. However, those works only considered the \ac{BEP} over a large number of subcarriers. It should be noted that the Gaussian approximation of the \ac{ICI} depends on the constellation order and the subcarrier considered. Thus we will show in this section results for the different constellation orders and subcarriers considered in previous sections, i.e., the subcarrier $150$ where the Gaussian approximation is better, and the subcarrier $300$ where the Gaussian approximation is the worst.

\begin{figure}[t]
	\centering
	\includegraphics[width=0.95\columnwidth]{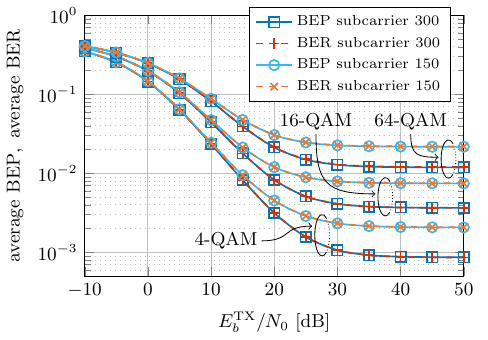}
	\caption{Average \ac{BEP} and average \ac{BER} versus $E_b^{\mathrm{TX}}/N_0$ for a symbol in subcarriers $150$ and $300$ considering $10^4$ realizations of the ITU-R vehicular channel model with $\nu_\mathrm{max}T = 0.05$.}
	\label{fig:bep_vs_ebno}
\end{figure}

\cref{fig:bep_vs_ebno} shows the average \ac{BEP} and \ac{BER} results versus the ratio of transmitted energy per bit to the noise power spectral density for $10^4$ realizations of the ITU-R vehicular channel model considering a single symbol for subcarriers $150$ and $300$, and a maximum normalized Doppler shift of $0.05$. The \ac{BER} is calculated using $10^4$ iterations (each iteration is specified at the beginning of \cref{sec:results}). \cref{fig:bep_vs_ebno} shows that both the analytical and the simulation results match for the three constellation orders and the two considered subcarriers for the entire range of $E_b^{\mathrm{TX}}/N_0$ values considered. As expected, the \ac{BEP} results obtained for subcarrier $300$ are lower than those of subcarrier $150$ because the \ac{ICI} affecting the border subcarriers is also lower.

\begin{figure}[t]
	\centering
	\includegraphics[width=0.95\columnwidth]{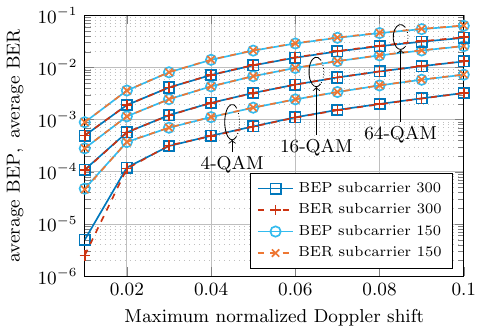}
	\caption{Average \ac{BEP} and average \ac{BER} versus maximum normalized Doppler shift for a symbol in subcarriers $150$ and $300$ considering $10^4$ realizations of the ITU-R vehicular channel model and $E_b^{\mathrm{TX}}/N_0 = 50$\,dB.}
	\label{fig:bep_vs_max_doppler}
\end{figure}

\cref{fig:bep_vs_max_doppler} shows the average \ac{BEP} and \ac{BER} versus the maximum normalized Doppler shift for $10^4$ realizations of the ITU-R vehicular channel model considering a single symbol for subcarriers $150$ and $300$, and $E_b^{\mathrm{TX}}/N_0 = 50$\,dB. As before, the \ac{BER} was calculated using $10^4$ iterations. Again, \cref{fig:bep_vs_max_doppler} shows that the analytical and the simulation results match for the three constellation orders and the two considered subcarriers for all the range of maximum normalized Doppler shifts considered. The only exception is the case of the minimum normalized Doppler shift considered ($0.01$) for subcarrier $300$ and $4$-QAM, where the \ac{BER} is slightly lower than the \ac{BEP}. 
In these results we also observe that, for each maximum normalized Doppler shift value, the \ac{BEP} results obtained for subcarrier $300$ are always lower than those of subcarrier $150$, because the \ac{ICI} is also lower.

Hence, the results show that, even for border subcarriers, the obtained \ac{BEP} values are a good approximation of the \ac{BER}. This can be justified because in the presence of Rayleigh fading the main contribution to the average \ac{BEP}, as can be seen in \cref{fig:bep_vs_symbol}, is due to a few symbols with a relatively low \ac{SINR}, and thus a high \ac{BEP}. As shown in \cref{sec:results_A}, we obtain good approximations of the \ac{BER} for deterministic channels and low \ac{SINR} values. Therefore when we consider the average of the \ac{BEP}, we obtain good approximations of the \ac{BER} because the main contribution to the average is due to symbols with low \ac{SINR} values. As shown in \cref{fig:bep_vs_max_doppler}, the results do not match only for the case of the border subcarrier with $4$-QAM and a low maximum Doppler shift value. This is because the symbols contributing to the average \ac{BER} have higher \ac{SINR} values where the \ac{BEP} approximation is worse as shown in \cref{sec:results_A}.

\section{Conclusions}

In this paper we proposed and studied an analytical approximation of the \ac{BEP} and provided a lower capacity bound for \ac{OFDM} systems when transmitting random Gray-coded \ac{QAM} symbols over deterministic doubly-selective channels. Assuming a general tapped delay channel model we firstly calculated the expressions for the channel response and \ac{ICI} coefficients affecting the \ac{OFDM} symbols and showed that the \ac{ICI} is not normally-distributed. However, considering deterministic channel realizations, i.e., that the channel response is previously known, we showed by means of Monte-Carlo simulations on three different standardized channel models (3GPP typical urban, 3GPP rural area, and ITU-R vehicular) that the \ac{ICI} is suitably approximated by a complex-valued normal distribution in realistic scenarios. The goodness of the approximation was found to be better as the constellation order increases, specially from $4$-\ac{QAM} to $16$-\ac{QAM}, whereas from $16$-QAM to $64$-QAM the improvement is not so large. It was also shown that for the border subcarriers (the ones closer to the edge of the effective used bandwidth) the normal approximation may be poor, but the goodness increases rapidly as we move away from the border subcarrier, reaching values near the maximum in just $4$ or $5$ subcarriers.

Next, we analyzed the capacity and the \ac{BEP} of the \ac{OFDM} systems considered. Firstly, we obtained an analytical expression for a lower bound of the capacity by means of the Shannon-Hartley theorem, assuming that the noise plus \ac{ICI} is Gaussian. Secondly, we showed that the \ac{BEP} can be approximated also by considering that the noise plus \ac{ICI} is Gaussian and using well-known \ac{BEP} formulas for Gray-coded \ac{QAM} symbols transmitted over \ac{AWGN} channels. To characterize the goodness of this approximation, we defined the metric of the error factor, expressed as the ratio of the \ac{BER} (which we calculated by means of Monte-Carlo simulations) to the \ac{BEP} (the proposed analytical approximation). Results of the instantaneous error factor metric showed that the instantaneous \ac{BEP} approximates well the instantaneous \ac{BER} for low \ac{SINR} values, whereas for larger \ac{SINR} values, the instantaneous \ac{BEP} was shown to be always lower than the instantaneous \ac{BER}, i.e., the \ac{BEP} is an upper bound of the \ac{BER}. Finally, we showed results of the averaged \ac{BEP} and \ac{BER} for several channel realizations versus the ratio of transmitted bit energy to noise spectral power density and versus the maximum normalized Doppler shift. Results were shown for two subcarriers, one at the middle and one on the edge where the normal approximation of the \ac{ICI} was shown to be worse. In these results the average \ac{BEP} matched almost exactly the average \ac{BER}, even for the border subcarrier. Thus, we conclude that the proposed \ac{BEP} approximation is indeed a good approximation of the average \ac{BER} in \ac{OFDM} systems.

The results obtained in this paper represent a great contribution to the research and industrial societies working on the performance evaluation of different deployments and network architectures for wireless communications in doubly-selective channels, which are a key scenario for \ac{5G} cellular systems. Different from previous related works in the literature, the performed study is not constrained to a specific kind of channel, but applicable to a general tapped delay one. Moreover, the proposed closed-form analytical expressions for the capacity and \ac{BEP} do not only allow for noticeably improving the efficiency of the evaluations (by discarding the need of computationally-costly Monte-Carlo simulations), but also provide a theoretical framework to optimize the parameters of the \ac{OFDM} communication systems directly impacting on the achievable performance. Finally, in order to help other researchers to easily use our results, we have made all the source code employed to obtain the results in this paper freely available.

\bibliographystyle{IEEEtran}
\bibliography{IEEEabrv,main}

\begin{thebibliography}{10}
\providecommand{\url}[1]{#1}
\csname url@samestyle\endcsname
\providecommand{\newblock}{\relax}
\providecommand{\bibinfo}[2]{#2}
\providecommand{\BIBentrySTDinterwordspacing}{\spaceskip=0pt\relax}
\providecommand{\BIBentryALTinterwordstretchfactor}{4}
\providecommand{\BIBentryALTinterwordspacing}{\spaceskip=\fontdimen2\font plus
\BIBentryALTinterwordstretchfactor\fontdimen3\font minus
  \fontdimen4\font\relax}
\providecommand{\BIBforeignlanguage}[2]{{%
\expandafter\ifx\csname l@#1\endcsname\relax
\typeout{** WARNING: IEEEtran.bst: No hyphenation pattern has been}%
\typeout{** loaded for the language `#1'. Using the pattern for}%
\typeout{** the default language instead.}%
\else
\language=\csname l@#1\endcsname
\fi
#2}}
\providecommand{\BIBdecl}{\relax}
\BIBdecl

\bibitem{cho2010mimo}
Y.~S. Cho, J.~Kim, W.~Y. Yang, and C.~G. Kang, \emph{MIMO-OFDM wireless
  communications with MATLAB}.\hskip 1em plus 0.5em minus 0.4em\relax John
  Wiley \& Sons, 2010.

\bibitem{ai2014challenges}
B.~{Ai}, X.~{Cheng}, T.~{Kürner}, Z.~{Zhong}, K.~{Guan}, R.~{He}, L.~{Xiong},
  D.~W. {Matolak}, D.~G. {Michelson}, and C.~{Briso-Rodriguez}, ``Challenges
  toward wireless communications for high-speed railway,'' \emph{{IEEE} Trans.
  Intell. Transp. Syst.}, vol.~15, no.~5, pp. 2143--2158, Oct 2014.

\bibitem{ArtigoMedidasTren_WCMC2017}
T.~Dom\'inguez-Bola{\~n}o, J.~Rodr\'iguez-{Pi\~neiro}, J.~A. Garc\'ia-Naya, and
  L.~Castedo, ``Experimental characterization and modeling of {LTE} wireless
  links in high-speed trains,'' \emph{Wireless Communications and Mobile
  Computing}, vol. 2017, no. 5079130, pp. 1--20, 2017, special Issue on
  Wireless Communications in Transportation Systems. Online access:
  \url{http://dx.doi.org/10.1155/2017/5079130}.

\bibitem{zhang2014measurement}
Y.~Zhang, Z.~He, W.~Zhang, L.~Xiao, and S.~Zhou, ``Measurement-based delay and
  doppler characterizations for high-speed railway hilly scenario,''
  \emph{International Journal of Antennas and Propagation}, vol. 2014, pp.
  1--8, 2014.

\bibitem{liu2012position}
L.~Liu, C.~Tao, J.~Qiu, H.~Chen, L.~Yu, W.~Dong, and Y.~Yuan, ``Position-based
  modeling for wireless channel on high-speed railway under a viaduct at 2.35
  {GHz},'' \emph{{IEEE} J. Sel. Areas Commun.}, vol.~30, no.~4, pp. 834--845,
  2012.

\bibitem{ArtigoPerdasPropagacion_Measurement2016}
L.~Zhang, J.~Rodr\'iguez-{Pi\~neiro}, J.~R. Fern\'andez, J.~A. Garc\'ia-Naya,
  D.~W. Matolak, C.~Briso, and L.~Castedo, ``Propagation modeling for
  outdoor-to-indoor and indoor-to-indoor wireless links in high-speed train,''
  \emph{Measurement}, vol. 110, pp. 43--52, June 2017, online access:
  \url{http://dx.doi.org/10.1016/j.measurement.2017.06.014}.

\bibitem{paier2009characterization}
A.~Paier, J.~Karedal, N.~Czink, C.~Dumard, T.~Zemen, F.~Tufvesson, A.~F.
  Molisch, and C.~F. Mecklenbr{\"a}uker, ``Characterization of
  vehicle-to-vehicle radio channels from measurements at 5.2 {GHz},''
  \emph{Wireless Personal Communications}, vol.~50, no.~1, pp. 19--32, Jul
  2009.

\bibitem{molisch2009survey}
A.~F. {Molisch}, F.~{Tufvesson}, J.~{Karedal}, and C.~F. {Mecklenbr{\"a}uker},
  ``A survey on vehicle-to-vehicle propagation channels,'' \emph{{IEEE}
  Wireless Commun.}, vol.~16, no.~6, pp. 12--22, Dec. 2009.

\bibitem{matolak2014modeling}
D.~W. {Matolak}, ``Modeling the vehicle-to-vehicle propagation channel: A
  review,'' \emph{Radio Science}, vol.~49, no.~9, pp. 721--736, Sep. 2014.

\bibitem{mecklenbrauker2011vehicular}
C.~F. {Mecklenbrauker}, A.~F. {Molisch}, J.~{Karedal}, F.~{Tufvesson},
  A.~{Paier}, L.~{Bernado}, T.~{Zemen}, O.~{Klemp}, and N.~{Czink}, ``Vehicular
  channel characterization and its implications for wireless system design and
  performance,'' \emph{Proc. IEEE}, vol.~99, no.~7, pp. 1189--1212, Jul. 2011.

\bibitem{viriyasitavat2015vehicular}
W.~{Viriyasitavat}, M.~{Boban}, H.~{Tsai}, and A.~{Vasilakos}, ``Vehicular
  communications: Survey and challenges of channel and propagation models,''
  \emph{{IEEE} Veh. Technol. Mag.}, vol.~10, no.~2, pp. 55--66, Jun. 2015.

\bibitem{walree2013propagation}
P.~A. {van Walree}, ``Propagation and scattering effects in underwater acoustic
  communication channels,'' \emph{{IEEE} J. Ocean. Eng.}, vol.~38, no.~4, pp.
  614--631, Oct 2013.

\bibitem{IMT-2020}
ITU-R, ``{ITU-R} {M}.2410-0 - {M}inimum requirements related to technical
  performance for {IMT}-2020 radio interface(s),'' Tech. Rep., Nov. 2017.

\bibitem{hasegawa2018high}
F.~Hasegawa, A.~Taira, G.~Noh, B.~Hui, H.~Nishimoto, A.~Okazaki, A.~Okamura,
  J.~Lee, and I.~Kim, ``High-speed train communications standardization in 3gpp
  5g nr,'' \emph{IEEE Communications Standards Magazine}, vol.~2, no.~1, pp.
  44--52, 2018.

\bibitem{li2018uav}
B.~Li, Z.~Fei, and Y.~Zhang, ``Uav communications for 5g and beyond: Recent
  advances and future trends,'' \emph{IEEE Internet of Things Journal}, vol.~6,
  no.~2, pp. 2241--2263, 2018.

\bibitem{3GPP_LTE_high_speed_enh2}
3GPP, ``Further performance enhancement for {LTE} in high speed scenario,''
  Tech. Rep., Jun. 2018, work item 800079.

\bibitem{3GPP_NR_HST}
------, ``{NR} support for high speed train scenario,'' Tech. Rep., Oct. 2019,
  work item 840092.

\bibitem{3GPP_EAV}
------, ``{5G} enhancement for {UAV}s,'' Tech. Rep., Jun. 2019, work item
  840083.

\bibitem{dang2020should}
S.~Dang, O.~Amin, B.~Shihada, and M.-S. Alouini, ``What should {6G} be?''
  \emph{Nature Electronics}, vol.~3, no.~1, pp. 20--29, 2020.

\bibitem{molisch2005wireless}
A.~F. Molisch, \emph{Wireless communications}.\hskip 1em plus 0.5em minus
  0.4em\relax John Wiley \& Sons, 2005.

\bibitem{russell1995interchannel}
M.~{Russell} and G.~L. {Stuber}, ``Interchannel interference analysis of {OFDM}
  in a mobile environment,'' in \emph{IEEE 45th Vehicular Technology
  Conference}, vol.~2, Jul. 1995, pp. 820--824.

\bibitem{wan2000bit}
L.~Wan and V.~Dubey, ``Bit error probability of {OFDM} system over frequency
  nonselective fast {R}ayleigh fading channels,'' \emph{Electronics Letters},
  vol.~36, pp. 1306--1307, Jul. 2000.

\bibitem{chiavaccini2000error}
E.~{Chiavaccini} and G.~M. {Vitetta}, ``Error performance of {OFDM} signaling
  over doubly-selective {R}ayleigh fading channels,'' \emph{IEEE Communications
  Letters}, vol.~4, no.~11, pp. 328--330, Nov. 2000.

\bibitem{nissel2017ofdm}
R.~Nissel and M.~Rupp, ``{OFDM} and {FBMC-OQAM} in doubly-selective channels:
  Calculating the bit error probability,'' \emph{{IEEE} Commun. Lett.},
  vol.~21, no.~6, pp. 1297--1300, 2017.

\bibitem{wang2006performance}
T.~Wang, J.~G. Proakis, E.~Masry, and J.~R. Zeidler, ``Performance degradation
  of {OFDM} systems due to {D}oppler spreading,'' \emph{{IEEE} Trans. Wireless
  Commun.}, vol.~5, no.~6, pp. 1422--1432, 2006.

\bibitem{pratschner2018versatile}
S.~Pratschner, B.~Tahir, L.~Marijanovic, M.~Mussbah, K.~Kirev, R.~Nissel,
  S.~Schwarz, and M.~Rupp, ``Versatile mobile communications simulation: the
  vienna {5G} link level simulator,'' \emph{EURASIP Journal on Wireless
  Communications and Networking}, vol. 2018, no. 226, 2018.

\bibitem{ArtigoSimuladorGTEC_IWSLS2016}
T.~Dom\'inguez-Bola{\~n}o, J.~Rodr\'iguez-{Pi\~neiro}, J.~A. Garc\'ia-Naya, and
  L.~Castedo, ``The {GTEC} 5{G} link-level simulator,'' in \emph{1st
  International Workshop on Link- and System Level Simulations (IWSLS2 2016)},
  Vienna, Austria, July 2016, online access:
  \url{http://dx.doi.org/10.1109/IWSLS.2016.7801585}.

\bibitem{nikaein2014openairinterface}
N.~Nikaein, M.~K. Marina, S.~Manickam, A.~Dawson, R.~Knopp, and C.~Bonnet,
  ``{OpenAirInterface}: A flexible platform for {5G} research,'' \emph{ACM
  SIGCOMM Computer Communication Review}, vol.~44, no.~5, pp. 33--38, 2014.

\bibitem{rupp2016vienna}
M.~Rupp, S.~Schwarz, and M.~Taranetz, \emph{The Vienna LTE-advanced simulators:
  Up and Downlink, Link and System Level Simulation}.\hskip 1em plus 0.5em
  minus 0.4em\relax Springer, 2016.

\bibitem{ns3}
\BIBentryALTinterwordspacing
``ns-3 network simulator.'' [Online]. Available: \url{https://www.nsnam.org/}
\BIBentrySTDinterwordspacing

\bibitem{mardia1970measures}
K.~V. Mardia, ``Measures of multivariate skewness and kurtosis with
  applications,'' \emph{Biometrika}, vol.~57, no.~3, pp. 519--530, 12 1970.

\bibitem{3GPPTR25.943}
3GPP, ``Universal {M}obile {T}elecommunications {S}ystem ({UMTS}); deployment
  aspects ({3GPP TR 25.943} release 15),'' Tech. Rep., Jul. 2018.

\bibitem{ITURM.1225}
ITU-R, ``Recommendation {ITU-R M.1225}, guidelines for evaluation of radio
  transmission technologies for {IMT}-2000,'' Tech. Rep., Feb. 1997.

\bibitem{xiao2006novel}
C.~Xiao, Y.~R. Zheng, and N.~C. Beaulieu, ``Novel sum-of-sinusoids simulation
  models for {R}ayleigh and {R}ician fading channels,'' \emph{{IEEE} Trans.
  Wireless Commun.}, vol.~5, no.~12, pp. 3667--3679, 2006.

\bibitem{IHARA197834}
S.~Ihara, ``On the capacity of channels with additive non-gaussian noise,''
  \emph{Information and Control}, vol.~37, no.~1, pp. 34--39, 1978.

\bibitem{883298}
D.~Yoon, K.~Cho, and J.~Lee, ``Bit error probability of {M}-ary quadrature
  amplitude modulation,'' in \emph{IEEE VTS Fall VTC2000 52nd Vehicular
  Technology Conference}, vol.~5, Sep. 2000, pp. 2422--2427.

\bibitem{Andrews97special}
L.~C. Andrews, \emph{Special Functions of Mathematics for Engineers},
  2nd~ed.\hskip 1em plus 0.5em minus 0.4em\relax SPIE Press, 1997, iSBN:
  9780819483713.

\bibitem{lapidoth1996nearest}
A.~Lapidoth, ``Nearest neighbor decoding for additive non-{G}aussian noise
  channels,'' \emph{{IEEE} Trans. Inf. Theory}, vol.~42, no.~5, pp. 1520--1529,
  1996.

\bibitem{bBootstrap_Efron2004}
B.~Efron and D.~V. Hinkley, \emph{An Introduction to the Bootstrap ({CRC}
  Monographs on Statistics \& Applied Probability)}, 1st~ed.\hskip 1em plus
  0.5em minus 0.4em\relax United States: Chapman \& Hall, 1994.

\end{thebibliography}

\begin{IEEEbiography}[{\includegraphics[width=1in,height=1.25in,clip,keepaspectratio]{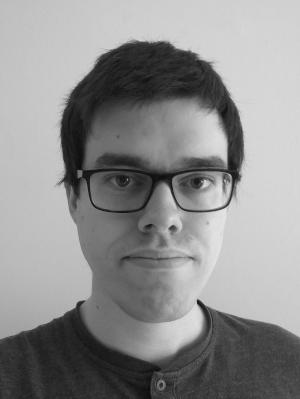}}
	]{Tomás Domínguez-Bolaño} received the B.S degree in Computer Engineering and the Ph.D. in Computer Engineering (with the distinction ``Doctor with European Mention'') from the University of A Coruña, A Coruña, Spain, in 2014 and 2018, respectively. Since 2014 he has been with the Group of Electronics Technology and Communications. In 2018 he was a Visiting Scholar with Tongji University, Shanghai, China. He is an author of more than 15 papers in peer-reviewed international journals and conferences. He was awarded with a predoctoral grant and two research-stay grants. His research interests include channel measurements, parameter estimation and modeling and experimental evaluation of wireless communication systems.
\end{IEEEbiography}

\begin{IEEEbiography}[
	{\includegraphics[width=1in,height=1.25in,clip,keepaspectratio]{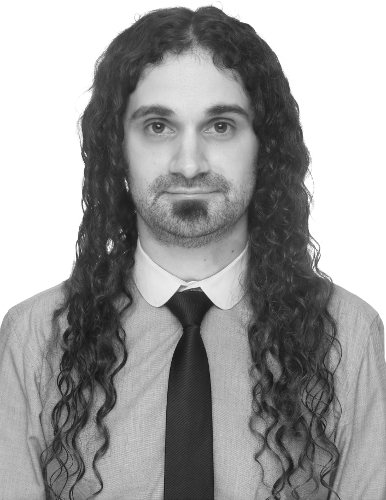}}
	]{José Rodríguez-Piñeiro} received the B.Sc. on  Telecommunications and the M.Sc. Degree in Signal Processing Applications for Communications from the University of Vigo (Pontevedra, Spain), in 2009 and 2011, respectively. Between June 2008 and July 2011, he was a researcher at the Department of Signal and Communications, University of Vigo (Pontevedra, Spain). From October 2011 he was a researcher at the Group of Electronics Technology and Communications of the University of A Coruña (UDC), obtaining his Ph.D. degree with the distinction “Doctor with European Mention” in 2016. After obtaining his Ph.D. degree with the he continued working as a Postdoctoral researcher at the same group until July 2017. On August 2017 he joined the College of Electronics and Information Engineering, Tongji University (P.R. China), becoming an Assistant Professor in 2020. From November 2012 he also collaborates with the Department of Power and Control Systems, National University of Asunción (Paraguay) in both teaching and research. He is the coauthor of more than 50 papers in peer-reviewed international journals and conferences. He is also a member of the research team in more than 25 research projects funded by public organizations and private companies. He was awarded with 6 predoctoral, postdoctoral and research stay grants. His research interests include experimental evaluation of digital mobile communications, especially for high mobility environments, including terrestrial and aerial vehicular scenarios.
\end{IEEEbiography}

\begin{IEEEbiography}[
	{\includegraphics[width=1in,height=1.25in,clip,keepaspectratio]{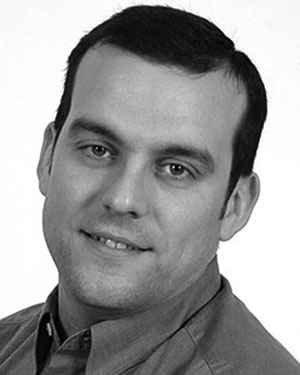}}
	]{José A. García-Naya} (S'07–M'10) received the M.Sc. and Ph.D. degrees in computer engineering from the University of A Coruña (UDC), A Coruña, Spain, in 2005 and 2010, respectively,where he has been with the Group of Electronics Technology and Communications, since 2005, and is currently an Associate Professor. He is the coauthor of more than 90 peer-reviewed papers in journals and conferences. His research interests include the experimental evaluation of wireless systems in realistic scenarios (indoors, outdoors, high mobility, and railway transportation), signal processing for wireless communications, wireless sensor networks, specially devoted to indoor positioning systems, and time-modulated antenna arrays applied to wireless communication systems. He is also a member of the research team of more than 40 research projects funded by public organizations and private companies.
\end{IEEEbiography}

\begin{IEEEbiography}[{\includegraphics[width=1in,height=1.25in,clip,keepaspectratio]{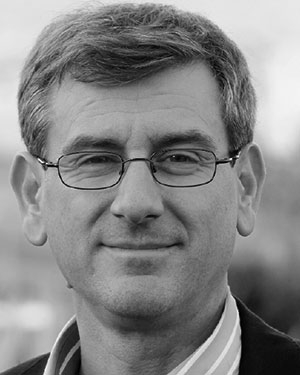}}]
	{ Luis Castedo } is currently Professor at the University of A Coruña (UDC), Spain. He received his PhD degree in Telecommunication Engineering from the Technical University of Madrid, Spain. Since 1994 he has been a Faculty Member with the Department of Computer Engineering, University of A Coruña, Spain, where he became Professor in 2001 and acted as Chairman between 2003 and 2009. He had previously held several research appointments at the University of Southern California (USC) and École supérieure d'électricité (SUPELEC). Between 2014 and 2018 he has been Manager of the Communications and Electronic Technologies (TEC) program in the State Research Agency of Spain. His research interests are signal processing, coding, hardware prototyping and experimental evaluation in wireless communications engineering.

	Prof. Castedo is coauthor of more than 300 papers in peer-reviewed international journals and conferences. He has also been principal investigator in more than 50 research projects funded by public organizations and private companies. He has advised 16 PhD dissertations. His papers have received three best student paper awards at the IEEE/ITG Workshop on Smart Antennas in 2007, at the IEEE International Workshop on Signal Processing Advances in Wireless Communications in 2013, and at the IEEE International Conference on Internet of Things (iThings) in 2017. He has been General Co-Chair of the 8th IEEE Sensor Array and Multichannel Signal Processing Workshop in 2014, and the 27th European Signal Processing Conference in 2019.
\end{IEEEbiography}

\end{document}